\documentclass[acmsmall]{acmart}
\AtBeginDocument{%
  }

\usepackage{pifont}
\usepackage{enumitem}
\usepackage{xcolor}
\usepackage{tikz}
\usetikzlibrary{positioning,arrows.meta,calc,decorations.pathreplacing}
\definecolor{revisionblue}{rgb}{0,0,0}
\newcommand{\revision}[1]{#1}
\usepackage[many]{tcolorbox}
\DeclareRobustCommand{\mybox}[2][gray!20]{%
\begin{tcolorbox}[
        breakable,
        left=0pt,
        right=0pt,
        top=0pt,
        bottom=0pt,
        colback=#1,
        colframe=#1,
        width=\linewidth, 
        enlarge left by=0mm,
        boxsep=5pt,
        arc=0pt,outer arc=0pt,
        ]
        #2
\end{tcolorbox}
}

\setcopyright{cc}
\setcctype{by}
\acmDOI{10.1145/3832133}
\acmYear{2026}
\acmJournal{PACMSE}
\acmVolume{3}
\acmNumber{ISSTA}
\acmArticle{ISSTA042}
\acmMonth{10}
\acmSubmissionID{issta26main-p366-p}
\received{2026-01-30}
\received[accepted]{2026-06-25}

\begin{document}

\title{Mind the Gap: An Empirical Study of Synchronization Gaps, Delays, and Missed Opportunities in Software Forks}

\author{Jiaying Zhu}
\orcid{0009-0007-4593-6207}
\email{jiaying007@e.ntu.edu.sg}
\affiliation{%
  \institution{Nanyang Technological University}
  \country{Singapore}}

\author{Lyuye Zhang}
\authornote{Corresponding author.}
\orcid{0000-0003-3087-9645}
\email{zh0004ye@e.ntu.edu.sg}
\affiliation{%
  \institution{Nanyang Technological University}
  \country{Singapore}}

\author{Jiahui Wu}
\orcid{0000-0001-6758-4635}
\email{jiahui004@e.ntu.edu.sg}
\affiliation{%
  \institution{Nanyang Technological University}
  \country{Singapore}}

\author{Chengyue Liu}
\orcid{0009-0003-7986-8004}
\email{CHENGYUE001@e.ntu.edu.sg}
\affiliation{%
  \institution{Nanyang Technological University}
  \country{Singapore}}

\author{Yang Liu}
\orcid{0000-0001-7300-9215}
\email{yangliu@ntu.edu.sg}
\affiliation{%
  \institution{Nanyang Technological University}
  \country{Singapore}}

\renewcommand{\shortauthors}{Zhu et al.}

\begin{abstract}
Fork-based development enables parallel evolution of software, but unsynchronized contributions create persistent divergence: security patches, bug fixes, and quality improvements often fail to propagate across fork families, leaving downstream users exposed to known vulnerabilities or bugs and missing massive opportunities to improve the other repositories in the family.
We present the first large-scale empirical study of fork synchronization, analyzing popular GitHub fork families with 3,820 actively maintained forks, and developed a monitoring platform to mine the valuable commits and promote their swift merging. 

Our findings reveal a synchronization paradox: while 90\% of submitted pull requests are merged, only 6.92\% of fork commits ever appear in PRs, leaving massive fork development permanently unsynchronized across the families.
Synchronization delay is pervasive and structurally uneven where fork propagation accounts for 72.9\% of end-to-end commit lifecycle delay.
Contrary to common assumptions, PR rejection is rarely caused by technical incorrectness; instead, 65\% of rejections stem from superseded contributions, process violations, or maintainer policy decisions.

Based on these insights, we develop a three-stage syncability assessment pipeline that identifies fork-local commits that are both \emph{sync-worthy} (broadly beneficial) and \emph{sync-eligible} (technically and policy-compatibly portable).
Applied to 0.5 million fork-local commits, our pipeline surfaces 12,284 sync-ready commit--repository pairs, demonstrating that our approach identifies practically valuable changes.
To further validate the security impact, we manually reviewed \textbf{153} security-related commit--repository pairs and confirmed \textbf{83} as potential 1-day vulnerabilities, for which we produced \textbf{35} proof-of-concept of exploit demonstrations and filed issues to the affected repositories.
Our monitoring platform enables continuous, near-real-time detection of synchronization opportunities across fork families, improving the sustainability of fork-based open-source ecosystems.

\end{abstract}

\begin{CCSXML}
<ccs2012>
   <concept>
       <concept_id>10011007.10011006.10011072</concept_id>
       <concept_desc>Software and its engineering~Software libraries and repositories</concept_desc>
       <concept_significance>500</concept_significance>
       </concept>
   <concept>
       <concept_id>10011007.10011006.10011073</concept_id>
       <concept_desc>Software and its engineering~Software maintenance tools</concept_desc>
       <concept_significance>500</concept_significance>
       </concept>
   <concept>
       <concept_id>10011007.10011006.10011071</concept_id>
       <concept_desc>Software and its engineering~Software configuration management and version control systems</concept_desc>
       <concept_significance>500</concept_significance>
       </concept>
 </ccs2012>
\end{CCSXML}

\ccsdesc[500]{Software and its engineering~Software libraries and repositories}
\ccsdesc[500]{Software and its engineering~Software maintenance tools}
\ccsdesc[500]{Software and its engineering~Software configuration management and version control systems}

\keywords{Software forks, fork synchronization, empirical study, pull requests, patch propagation, one-day vulnerabilities, mining software repositories}

\maketitle

\section{Introduction}
\label{sec:intro}

Fork-based development enables parallel evolution~\cite{dabbish2012social}: forks can rapidly prototype features, adapt to local constraints, and explore alternative designs while still benefiting from a shared upstream codebase. 
Multiple forks along with the original repository can form a fork family that retains substantial code overlap by sharing a common code base.
However, parallelism also introduces a fundamental drawback, namely, unsynchronized contributions. When useful changes fail to propagate within a fork family, projects diverge in behavior and quality and may retain known defects or vulnerabilities long after they are addressed upstream.
For example, Vim~\cite{vim} and Neovim~\cite{neovim} (a hard fork~\cite{pan2024automating} of Vim) have repeatedly shared code paths that led to command execution vulnerabilities: CVE-2019-12735~\cite{nvd2019_12735} affected both Vim~\cite{vim_repo} and Neovim~\cite{neovim_repo}, requiring coordinated fixes across the fork family, illustrating how downstream forks can remain exposed until patches are ported and released. 
Beyond security patches, failure to synchronize bug fixes and performance improvements can similarly compromise the quality and reliability of other repositories within the fork family.

This parallel-development setting is prone to inefficiencies~\cite{zhou2020forks}, including lost or duplicate contributions and fragmented communities, making cross-repository synchronization a recurring challenge rather than an exceptional event~\cite{zhou2019whatthefork}. The stakes are especially high for security: \textit{one-day vulnerabilities} can persist in forks after the origin is patched because fixes do not reliably propagate at the commit level~\cite{lefeuvre2025oneday,pan2024automating}. Accordingly, existing work has largely studied synchronization through a security-centric lens (e.g., accelerating security fix propagation~\cite{machiry2020spider} and measuring patch diffusion delays in downstream ecosystems~\cite{zhang2021android,yi2023blockscope,andreina2023patchprop}). However, it remains unclear how broadly useful general changes (bug fixes, quality improvements, tests/docs) synchronize at scale and how synchronization lifecycle and lag shape what ultimately propagates, let alone how to automatically facilitate the synchronization among fork families regarding all useful changes.

To bridge this gap, we conduct the first large-scale empirical study to characterize synchronization in popular GitHub repositories by explicitly identifying fork families and quantifying their unsynchronized contributions. \revision{A central motivation is that synchronization gaps are typically invisible to downstream users: a fork may have been actively maintained at adoption time and only later stopped tracking upstream fixes, leaving users exposed without any warning signal. Our goal is therefore visibility into these latent gaps, not an assumption that all forks remain maintained indefinitely.} Building on this evidence, we move beyond problem identification to ask how synchronization can be facilitated in a manner that is accurate, automated, and timely. Our final deliverable is a continuous, near-real-time platform that detects syncable commits and operationalizes their propagation (e.g., via actionable recommendations or PR creation), thereby improving the sustainability of fork-based open-source communities.

Achieving this goal requires capabilities that prior work has not yet delivered. First, existing studies often focus on domain-specific ecosystems (e.g., Linux vendors or blockchain forks), whereas we target general-purpose GitHub repositories at scale. Second, a fine-grained understanding of already synchronized contributions at the commit level—particularly their intent and timeliness—is limited, leaving the lifecycle of synchronization insufficiently explained. Third, the reasons why synchronization attempts fail have not been systematically studied: rejected fork$\rightarrow$origin PRs and their rejection rationales remain under-explored. Finally, no prior approach provides a real-time detector for syncable commits using fine-grained criteria that jointly consider technical feasibility, repository policies, and practical usefulness, which are all necessary to avoid noisy or low-value synchronization attempts.

We addressed these gaps via four research questions and built an end-to-end monitoring platform based on the findings. 
RQ1 characterizes synchronization pathways and reveals the sychronization gaps among the repositories in the fork family.
RQ2 quantifies synchronization lag in both fork$\rightarrow$origin and origin$\rightarrow$fork directions, clarifying where delay accumulates and measures the propagation life cycle of commits from creation to family-level synchronization; RQ3 analyzes what kinds of contributions synchronize, and why some are rejected, empirically yielding generally beneficial types;  and RQ4 shifts to fork-local native commits, evaluating which changes are both \textit{sync-worthy} and \textit{sync-eligible} yet remain unsynchronized to promote effective synchronization. Building on these results, we develop a monitoring platform that tracks syncable commits across fork families, assesses their synchronizability conservatively, and can automatically create PRs to facilitate safe and low-friction synchronization.
To validate the security impact, we manually reviewed 153 security-related commit--repository pairs and confirmed 83 as potential 1-day vulnerabilities where the target repository had not applied the fix, producing 35 proof-of-concept demonstrations and filing issues to draw maintainer attention.
Our contributions are threefold:

\begin{itemize}[leftmargin=9pt]
    \item We are the first to quantify fork-family synchronization as a staged process that jointly models exposure, rejection rationales, and bidirectional lag, and then operationalize commit-level detection with policy/usefulness screens.
    \item We reveal what kinds of changes are successfully synchronized and why synchronization attempts fail, thereby providing actionable insights on how synchronization can be facilitated.
    \item We develop a monitoring platform that continuously tracks sync-ready commits in near real time and evaluates their synchronization potential using fine-grained criteria that jointly consider technical feasibility, practical usefulness, and policy compatibility with target repositories. We further validate the security impact by manually confirming 83 potential 1-day vulnerabilities out of 153 security-related pairs, producing 35 proof-of-concept demonstrations.
\end{itemize}

\section{Background and Motivating Example}

\subsection{Background}
\label{sec:background}
Modern open-source development frequently involves parallel development across an origin repository and a large number of its forks. In the GitHub ecosystem, a fork is initially created as an exact copy of its origin repository, but it may evolve in one of two very different directions~\cite{zhou2019whatthefork}. Some forks are short-lived and exist only to stage a pull request (PR)~\cite{gousios2014exploratory}. Others continue to develop independently, introducing new features, refactoring, or patches that never return to the origin. These ``hard forks'' often emerge when contributors face obstacles such as unresponsive maintainers or rejected pull requests~\cite{zhou2020forks}. Conversely, origin maintainers also introduce updates, such as bug fixes or security patches, that may never reach the forks. This bidirectional flow of commits creates a poorly understood ecosystem where important changes may fail to propagate.

In this ecosystem, we consider three primary entities:
\begin{itemize}[leftmargin=12pt]
    \item \textbf{Origin repository}: the canonical upstream project from which the forks are created.
    \item \textbf{Forks}: developer- or organization-owned copies. \revision{A fork may be a short-lived PR-staging copy, a passive mirror, or an independently evolving code line. Our study focuses on independently evolving forks with native post-fork commits, since these forks can diverge from the origin and create synchronization gaps.}
    \item \textbf{Commit flows}: bidirectional synchronizations between origin and forks, either through PRs, cherry-picks~\cite{bunyakiati2017cherry}, or manual porting.
\end{itemize}

An ideal ecosystem maintains a healthy synchronization: forks contribute improvements upstream, and upstream fixes propagate downstream. However, in practice, many commits, especially critical ones—remain unsynchronized. As forks and origins continue diverging, incompatibilities increase, and important updates may be silently missed.

\begin{figure}[t!]
	\centering
\includegraphics[width=0.9\linewidth]{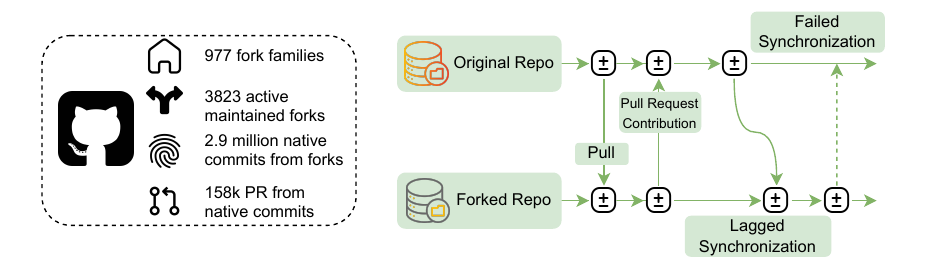}
	\caption{Scenario Illustration of Commits Synchronization Between the Origin and Fork}
	\label{fig:background}
\end{figure}

When commits fail to propagate, projects accumulate the gaps of changes: the set of commits that should have been adopted by another repository but were not. The gap is particularly harmful when the missing commits include security patches, correctness fixes, or compatibility adjustments. Downstream users of stale forks may unknowingly rely on vulnerable or incorrect code, while upstream maintainers may miss improvements already implemented in the forks.

\subsection{Motivating Example}
TARmageddon (CVE-2025-62518, CVSS 8.1)~\cite{nvd2025_62518} is a boundary-parsing flaw in the async Rust tar ecosystem that enables archive-entry smuggling and can lead to remote code execution in downstream applications that unpack untrusted tar files~\cite{hackernews2025_tarmageddon}.
Crucially, the vulnerability spans a fork family rooted at \texttt{async-tar}~\cite{asynctar2025repo} (a Rust library for asynchronous tar archive I/O) and includes its derivatives. \revision{The most widely depended-on fork, \texttt{tokio-tar}~\cite{tokiotar2025repo}, was effectively abandoned and would not receive a patch, forcing downstream users to migrate to a maintained fork, \texttt{astral-tokio-tar}~\cite{astraltokiotar2025repo} (a recent fork maintained by the Astral team, fixed in v0.5.6), rather than simply ``pulling'' upstream fixes~\cite{heise2025_tarmageddon,edera2025_tarmageddon}.
This incident illustrates the core risk we study: even when a fix exists within the fork family, unsynchronized or unmaintained forks can strand large downstream populations on vulnerable code. The fundamental difficulty is visibility: tokio-tar's drift from upstream was not announced or signposted to its downstream users; many continued depending on it after maintenance had silently lapsed. Our goal is therefore to surface such latent synchronization gaps as early as possible so that downstream users and maintainers can make informed decisions, rather than to assume that all forks remain maintained indefinitely.}

These observations motivate a systematic investigation into how often forks evolve independently, how contributions flow between origins and forks, what types of improvements arise in this parallel development, and how much technical lag accumulates across the ecosystem.

\section{Data Collection}

To construct a representative dataset for analyzing the technical lag phenomenon in fork ecosystems, we first collected the top open-source repositories from GitHub ranked by the number of stars. To ensure the dataset focuses on software projects, we filtered out non-code repositories using keyword-based heuristics (e.g., excluding dataset mirrors, tutorials, and pure documentation projects). After filtering, we retained 2,000 origin repositories. These popular projects typically exhibit active maintenance and extensive forking activity, making them suitable for studying contribution flows between origins and forks.

\revision{\textbf{Sample-selection rationale.} We select popular origins, ranked by stars, for two reasons. First, this choice follows prior large-scale GitHub fork studies~\cite{zhou2019whatthefork, zhou2020forks}, which also focus on repositories with substantial public activity. Second, popular repositories are more likely to have active fork ecosystems and established contribution workflows, making cross-repository synchronization both observable and practically relevant. At the same time, star-based selection can introduce popularity bias, as noted in OSS sampling and ranking studies~\cite{malviya2024role, borges2016understanding, borges2018githubstars, munaiah2017curating}. We therefore make this criterion explicit and account for its implications in the threats-to-validity discussion. The resulting scale of 2,000 origins is comparable to or larger than prior large-scale fork studies~\cite{zhou2019whatthefork, zhou2020forks}, which typically analyze hundreds to low thousands of repositories, while remaining feasible for exhaustive downstream fork collection and LLM-assisted labeling.

This sample yields 3,820 actively maintained forks and 2.9~million post-fork commits, which is sufficient to expose long-tailed ecosystem behavior without sub-sampling. We also considered initial selection by fork-graph complexity, such as fork count or fork-network depth, but did not use it as the sampling criterion. Fork-graph complexity is a downstream property of an origin: it can only be measured after retrieving and traversing the full fork network for each candidate, so it cannot guide the initial selection without first running the same collection pipeline. In our final sample, the 2,000 star-ranked origins yield 5.8~million forks (mean 2,901 forks per origin) and an average of 3.9 actively maintained forks per origin, confirming that this end of the distribution is well represented.}

We first examined the prevalence of parallel development among all forks identified in our dataset.
Across the 2,000 origin repositories, we collected 5,801,648 forks in total.
From each selected repository, we retrieved all publicly available forks via the GitHub REST and GraphQL APIs~\cite{kalliamvakou2014perils,gousios2012ghtorrent}. To ensure the analysis focused on software projects rather than documentation or configuration repositories, we heuristically filtered out non-code repositories by keywords such as dataset mirrors, tutorials, and pure documentation projects. After this filtering step, we obtained a total of \textbf{2,900,824} forks.
We focus on direct forks of each origin rather than nested forks (forks of forks). This choice is well-supported: GitHub's fork network architecture links all forks to a shared root repository, and the standard pull request workflow directs contributions to the upstream origin rather than intermediate forks~\cite{jiang2017forks}. Prior empirical studies of fork ecosystems similarly analyze direct origin--fork relationships, as nested forking is rare in practice and typically reflects transient experimentation rather than sustained parallel development~\cite{zhou2019whatthefork}.

As we aim to study the synchronization among origin repository and active forks, it is necessary to avoid analyzing the temporary forks that are only used to create PRs to the origin repository. To achieve this, we applied filters to only count those still under active maintenance. 
\revision{We applied three filtering criteria to identify actively and independently developed forks. 
\ding{172} The repository must have been updated with new commits within the 12~months preceding our data-collection snapshot in 2025, ensuring that it is currently maintained.
\ding{173} The repository must have at least ten stars; prior work has shown that temporary or PR-only forks typically attract very few stars, whereas star count serves as a reasonable proxy for project popularity and sustained usage~\cite{malviya2024role}. 
\ding{174} The repository must contain at least 100 commits introduced after forking that do not originate from the upstream project, following the criteria from~\cite{zhou2019whatthefork}. These commits are named as \texttt{native commits}.
This criterion ensures that the fork exhibits independent parallel development, as a single commit may correspond to an isolated contribution, while multiple new commits indicate sustained development beyond the original codebase.}

After applying the criteria, we retained \textbf{3,820} active forks, representing only \textbf{0.08\%} of all forks. Due to the removal of forks, the final number of origin-fork clusters dropped to 977.
On average, each origin repository had \textbf{3.9} actively maintained forks.
These results demonstrate that while the vast majority of forks on GitHub are temporary or inactive, a non-negligible subset exhibits sustained, independent development and thus has significant potential to diverge from their origins. \revision{The forks removed by these criteria are predominantly PR-only social forks, mirrors, and lightly modified tracking copies whose history is dominated by upstream pulls rather than native commits. They are out of scope because our study targets synchronization gaps arising from parallel evolution rather than passive upstream tracking, and they are therefore excluded from the statistics reported throughout the paper.}

For each remaining origin–fork cluster, we then collected all commits and PRs. Specifically, we extracted (1) all merged PRs submitted to the 977 origin repositories (3.0 million in total) and (2) all commits from both the origin and its active forks. In total, this process yielded 25 million commits in the origin repositories and 2.9 million unique commits across the corresponding forks. There were 158,482 PRs created from the forks to the origin repositories that were derived from unique commits. These data form the empirical foundation for our subsequent analysis of contribution flows and technical lag within the fork ecosystem.

\section{Empirical Study}

\subsection{Research Questions}

This section organizes our empirical analysis around the paper's overarching objective: facilitating the synchronization of useful changes within fork families so that parallel development remains sustainable rather than fragmenting into long-term divergence. First, we establish the current synchronization status of fork families and quantify the gap between fork-local development and cross-repository integration (\textbf{RQ1}). Second, recognizing that synchronization is only beneficial if it is timely, we measure how delay accumulates along the synchronization lifecycle in both fork$\rightarrow$origin and origin$\rightarrow$fork directions (\textbf{RQ2}). Third, to understand how synchronization can be improved, we analyze what kinds of changes successfully synchronize and why attempts fail, by characterizing the intentions of merged contributions and the rejection rationales of failed ones (\textbf{RQ3}). Finally, building on these insights, we operationalize a monitoring platform that continuously surfaces sync-ready fork-local commits and evaluates their synchronizability under conservative criteria (\textbf{RQ4}).

\begin{figure}[t!]
	\centering
\includegraphics[width=0.9\linewidth]{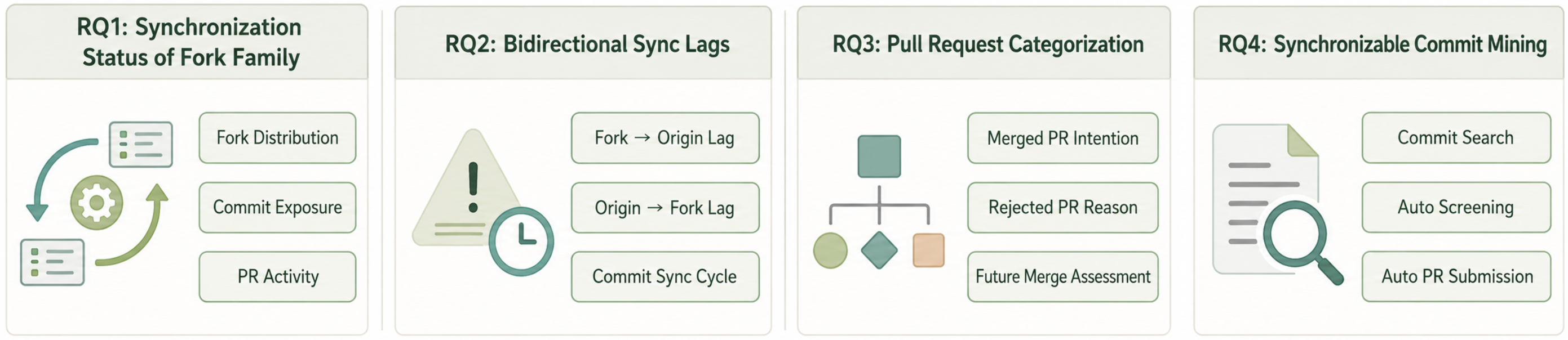}
	\caption{Overview of Research Questions}
	\label{fig:overview}
\end{figure}

\begin{itemize}[leftmargin=9pt]
    \item \textbf{RQ1: Synchronization status.}
    How much development in fork families is synchronized?

    \item \textbf{RQ2: Synchronization timeliness.}
    When synchronization occurs, how long does it take, and where does the delay accumulate over the lifecycle?

    \item \textbf{RQ3: What syncs and what fails.}
    What kinds of changes are merged, and why do synchronization attempts get rejected?

    \item \textbf{RQ4: Missed synchronization opportunities.}
    Which fork-local unique commits are both sync-worthy and sync-eligible yet remain unsynchronized, and what opportunities do they represent for improved collaboration?
\end{itemize}

\subsection{RQ1: Prevalence of Parallel Development}

\subsubsection{RQ1.1: Fork Distribution and Concentration}

\begin{figure}[t]
  \centering
  \includegraphics[width=\linewidth]{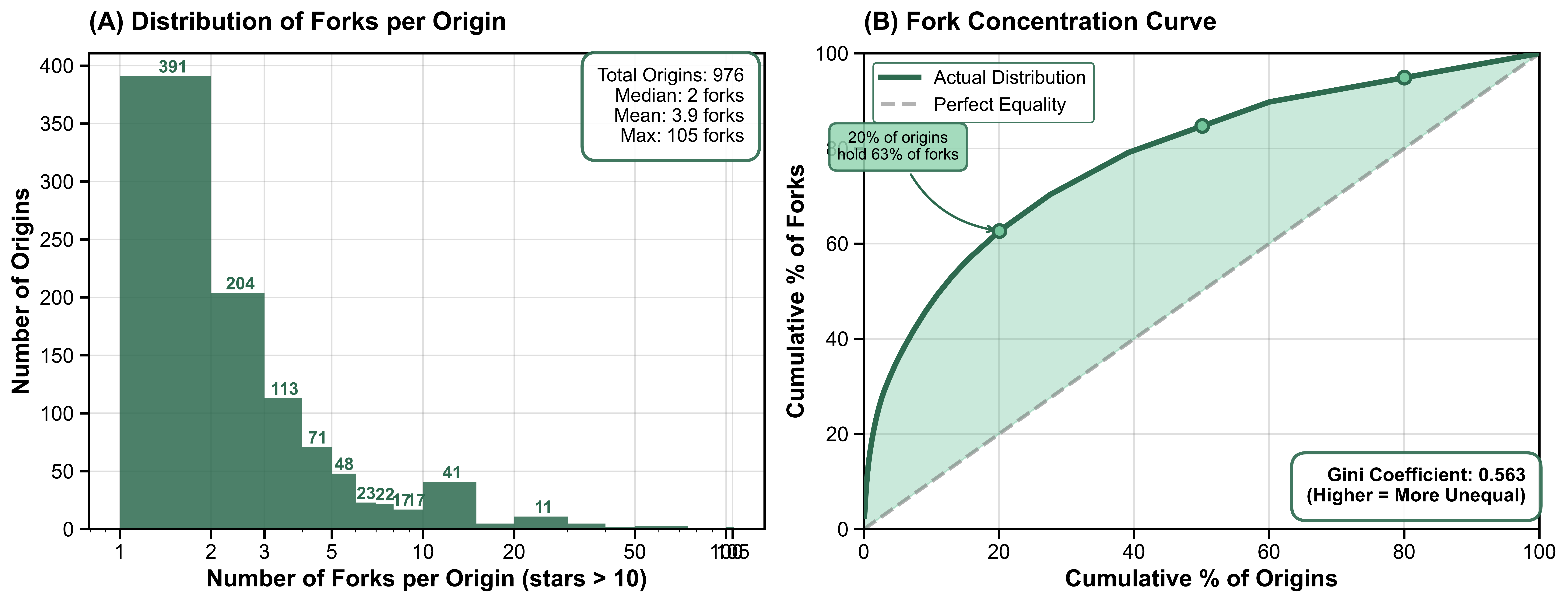}
  \caption{Fork distribution and concentration across origin repositories.
  (A) Distribution of the number of active forks per origin (log-scale on the x-axis).
  (B) Fork concentration curve (Lorenz-style), showing strong inequality in fork activity.
  }
  \label{fig:fork-distribution}
\end{figure}

\revision{\textbf{Fork activity is concentrated in a small subset of origins.}
Across 977 origin repositories, we identified 3,820 active forks, with 1--207 forks per origin. Figure~\ref{fig:fork-distribution}B quantifies the imbalance using a Lorenz-style concentration curve: the top \textbf{20\%} of origins account for approximately \textbf{63\%} of all active forks, yielding a Gini coefficient of \textbf{0.563}. Figure~\ref{fig:fork-distribution}A shows the corresponding long-tailed distribution from the per-origin angle: most origins host one or two active forks, while a small subset hosts dozens. Intensive fork-based development is therefore a property of a small fork-dense subset of origins rather than of the ecosystem at large.

This concentration has direct consequences for synchronization. The coordination workload within a fork family scales with its size: more parallel development branches mean more redundant fixes and more opportunities for commit staleness. In the fork-dense subset (e.g., Linux kernel variants and OpenSSL distributions), one well-propagated fix can reach hundreds of downstream repositories, whereas the origin with one or two forks faces a much lighter burden. Ecosystem-wide averages therefore understate the workload. We accordingly conduct our subsequent analyses at the per-family level and direct monitoring and tooling effort to the fork-dense families.}

\mybox{
\noindent\revision{\textbf{Finding~1:}
Fork activity is concentrated in a small fork-dense subset of origins: the top 20\% of origins account for over 60\% of all active forks (Gini = 0.56), with a median of \textbf{8} active forks per origin in the top quintile (maximum = \textbf{105}). Synchronization pressure scales with family size and is amplified in this dense subset.}
}

\subsubsection{RQ1.2: The Synchronization Paradox---High PR Activity, Low Commit Exposure}
\label{sec:rq1-synchronization-paradox}

\revision{RQ1.2 is scoped to fork$\to$origin exposure because PR submission is the canonical observable pathway for forks to expose changes upstream; origin$\to$fork synchronization proceeds through pulling, merging, or manual porting rather than PRs, so a symmetric PR-exposure metric is not meaningful in that direction. We therefore measure exposure at one direction here in RQ1 and revisit both directions at the lag level in RQ2, where shared-commit history signals allow direct comparison.}

At first glance, PR-based synchronization appears healthy: across our dataset, forks submit pull requests at scale, with the 2,000 origin repositories receiving over 3 million PRs in aggregate.
These submitted PRs enjoy a high merge success rate of \textbf{90.11\%}, suggesting that maintainers are receptive to fork contributions.
However, this apparent success masks a deeper problem: PR-level metrics capture the success of synchronization attempts once they are made, but they do not reflect how many native commits are covered in the PR events.

\textit{Measuring commit exposure.}
To quantify actual synchronization coverage, we define the \textbf{Commit Exposure Ratio (CER)} as the fraction of post-fork commits that appear in at least one pull request submitted to the origin:
\[ \text{CER} = \frac{\#\text{post-fork commits appearing in PRs}}{\#\text{post-fork commits}}. \]

\revision{\textbf{Counting commits toward CER: original commits and history-rewriting-aware commits.} The CER counts a post-fork commit as exposed only when its exact SHA appears in the commit metadata of at least one fork$\to$origin PR\@. However, this measure may miss the SHA-changed commit when a contributor rebases or amends the commit before submission, since rebase and amend produce a different SHA for content-equivalent code. 

To capture such re-submissions we extend the CER with a three-step classification pipeline. For each commit in a fork$\to$origin PR we (1) check whether its SHA appears in the fork's commit set and label it \emph{original} if so; (2) otherwise compare its diff against every fork unique commit (a commit in the fork but not in the origin) and label it \emph{derivative} if the diff similarity is at least 85\% or one diff's line set is a subset of the other's; (3) otherwise check whether its SHA appears in the origin and label it \emph{from-origin}. Diff similarity is the Jaccard similarity of the added + removed line sets of the two unified diffs, after normalizing whitespace and stripping diff metadata (file headers, hunk markers, and line numbers); the 85\% threshold and the subset rule are evaluated on the same line-set representation. 

The history-rewriting-aware CER counts both \emph{original} and \emph{derivative} commits as exposed, while the original CER counts only \emph{original}. An exact-SHA-only baseline yields a more conservative median CER of 6.92\% (mean 18.38\%); the rewriting-aware variant we adopt as the canonical CER for the rest of this paper raises these to 7.19\% and 22.78\% respectively.}

Despite the high PR merge rate, commit exposure is therefore strikingly low: the median fork exposes only \revision{\textbf{7.19\%}} of its post-fork commits through PRs (mean: \revision{22.78\%}).
In other words, for a typical fork, more than 90\% of commits never enter the upstream review process.
This low exposure is pervasive: \textbf{66.17\%} of all fork commits never appear in any PR, and in \textbf{86.2\%} of fork--origin pairs, more than half of fork commits remain fork-local.

\begin{figure}[t]
    \centering
    \includegraphics[width=0.7\linewidth]{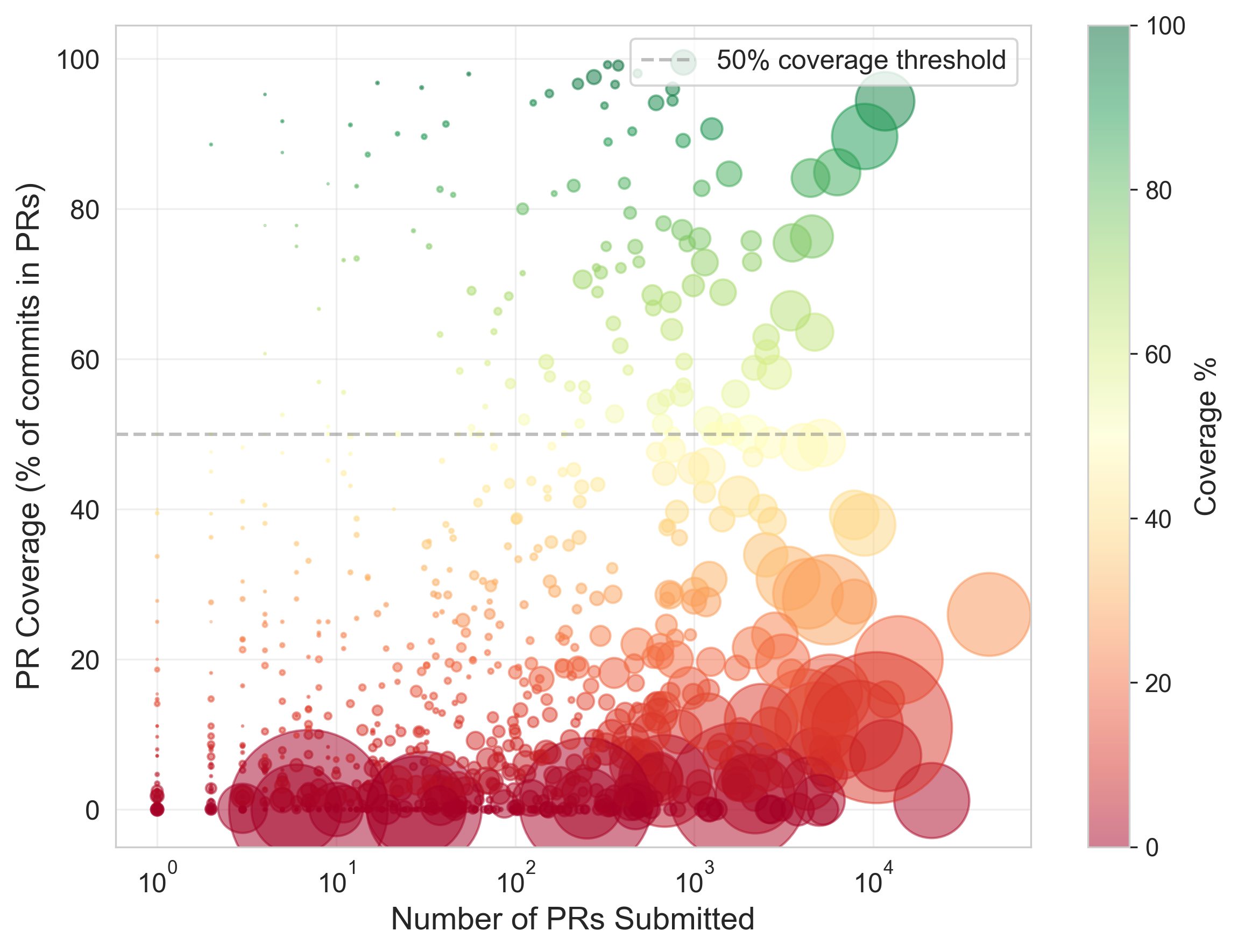}
    \caption{The synchronization paradox: PR activity versus commit-level coverage. Each point represents a fork--origin pair. 
    Bubble size indicates the total number of fork commits. 
    }
    \label{fig:synchronization-paradox}
\end{figure}

Figure~\ref{fig:synchronization-paradox} visualizes this disconnect: although many forks submit hundreds of PRs, the majority lie well below 50\% commit coverage (75\% of pairs have CER below 27.59\%).
This paradox persists regardless of project popularity---while popular origins attract more PR activity ($\rho = 0.29$), popularity shows almost no correlation with CER ($\rho = 0.07$).
Among origins with over 10K stars, 51\% have CER below 10\%, confirming that ecosystem prominence drives contribution attempts but not comprehensive synchronization.

In absolute terms, nearly \textbf{3 million commits} across the dataset remain fork-local, never appearing in any PR to the origin.
We admit that many such commits are intentionally fork-specific, reflecting customization, experimentation, or divergent project goals.
Nevertheless, given the sheer volume of post-fork development, even a small fraction of sync-worthy changes within this large fork-local pool would represent substantial missed synchronization opportunities.

\mybox{\textbf{Finding 2:} 
While submitted PRs are merged at a high rate (90\%), most fork development never reaches the PR stage at all, the median fork exposes only \revision{7.19\%} of its commits.
This selective exposure raises critical concerns: unsynchronized commits may include security patches, bug fixes, or valuable features that remain siloed in individual forks, potentially leaving downstream users exposed to known vulnerabilities.
}

\subsection{RQ2: Synchronization Lag and Temporal Fragmentation}
\label{sec:rq2-lag}

\revision{RQ1 reported a high merge rate (\textbf{90}\%) for submitted PRs, but acceptance does not imply timeliness. Substantial delays may arise even for PRs that are ultimately merged, and PR review covers only one of three temporal stages in the end-to-end synchronization lifecycle of a commit. RQ2 therefore shifts the focus from whether fork-local commits reach the origin to when they do, in which stage of the lifecycle delay accumulates, and how long an integrated change takes to propagate back across the fork family.}

\revision{We model fork-family synchronization as a three-phase pipeline spanning two units of observation as in~Figure~\ref{fig:lag-timeline}. Two of the phases characterize the fork$\to$origin pathway and are defined entirely with respect to a single PR object: all timestamps in these definitions originate from the same PR, and the two phases sum to the per-PR fork$\to$origin lag. The third phase characterizes another direction, the origin$\to$fork pathway, in which synchronization occurs through fork-side pulling rather than through PR submission; its unit of observation is therefore a commit-fork pair.}

\revision{%
\begin{itemize}[leftmargin=*]
    \item \textbf{Fork$\to$Origin synchronization} is observable at the PR level, since the canonical contribution channel from a fork to its origin is the submission of a PR for review and merging by the origin's maintainers. The lag of an individual PR decomposes into two sequential phases, each defined with respect to events (e.g., creation, merging) of the same PR (see Fig.~\ref{fig:lag-timeline}):
    \begin{itemize}[leftmargin=*]
        \item \textbf{Developer Batching Phase.} For a given PR, the elapsed time between the authoring timestamp of the last fork commit and the creation timestamp of that PR. This phase quantifies developer-side batching and local iteration before upstream submission; batching lag is determined per the criterion under \emph{Lag Determination} below.
        \item \textbf{Maintainer Review Phase.} For the same PR, the elapsed time between the PR's creation timestamp and its closure timestamp (whether merged or closed without merging). This phase quantifies maintainer-side review and decision latency; review lag is determined by the same criterion. Both phases are defined over the same PR, so their durations sum to the per-PR fork$\to$origin lag.
    \end{itemize}

    \item \textbf{Origin$\to$Fork synchronization}, by contrast, is realized when a fork pulls commits from the origin rather than through PR submission. \revision{Here ``pulling'' refers to the fork-sync operations documented by GitHub~\cite{githubdocsforksync} (``Sync fork'' in the web UI and the equivalent \texttt{git fetch}/\texttt{merge}/\texttt{rebase} from \texttt{upstream}). Other forms of propagation, such as manual porting, reimplementation, or propagation through an intermediate repository, cannot be identified reliably from public Git/GitHub metadata and are outside this measurement.} The corresponding lag is captured as the \textbf{Fork Pulling Phase}, defined as the elapsed time between the authoring timestamp of an upstream commit in the origin and the timestamp at which a given fork incorporates that commit into its own history. Since pulling lacks an associated PR object, the unit of observation for this phase is a commit-fork pair; pulling lag is determined by the same criterion.
\end{itemize}
}%

\revision{%
\begin{figure}[t]
\centering
\begin{tikzpicture}[
    scale=0.85, transform shape,
    >={Stealth[length=2mm]},
    every node/.style={font=\small, color=revisionblue},
    eventdot/.style={circle, fill=revisionblue, inner sep=1.5pt},
    phaselabel/.style={font=\small\itshape, align=center, color=revisionblue},
    dirlabel/.style={font=\scriptsize, color=revisionblue},
    lanelabel/.style={font=\small\itshape, color=revisionblue, anchor=east},
    rolelabel/.style={font=\scriptsize\itshape, color=revisionblue, align=center},
    flowarrow/.style={->, thick, color=revisionblue, dashed}
]
\draw[decorate, decoration={brace, amplitude=4pt}, thick, color=revisionblue]
    (0.6,2.55) -- (4.7,2.55)
    node[midway, above=4pt, phaselabel] {Developer\\Batching Phase};
\draw[decorate, decoration={brace, amplitude=4pt}, thick, color=revisionblue]
    (4.7,2.55) -- (8.8,2.55)
    node[midway, above=4pt, phaselabel] {Maintainer\\Review Phase};
\draw[decorate, decoration={brace, amplitude=4pt}, thick, color=revisionblue]
    (8.8,2.55) -- (12.9,2.55)
    node[midway, above=4pt, phaselabel] {Fork\\Pulling Phase};

\draw[->, thick, color=revisionblue] (0,2.0) -- (14.0,2.0);
\node[anchor=west, font=\scriptsize, color=revisionblue] at (14.0,2.0) {time};
\draw[thick, color=revisionblue] (0.6,2.15)  -- (0.6,1.85);
\draw[thick, color=revisionblue] (4.7,2.15)  -- (4.7,1.85);
\draw[thick, color=revisionblue] (8.8,2.15)  -- (8.8,1.85);
\draw[thick, color=revisionblue] (12.9,2.15) -- (12.9,1.85);

\draw[color=revisionblue!40, thin] (-0.1,0.6) -- (14.0,0.6);
\node[lanelabel] at (-0.2,0.6) {Fork-side};
\draw[color=revisionblue!40, thin] (-0.1,-0.9) -- (14.0,-0.9);
\node[lanelabel] at (-0.2,-0.9) {Origin-side};

\draw[dotted, color=revisionblue, thin] (0.6,1.85)  -- (0.6,0.6);
\draw[dotted, color=revisionblue, thin] (4.7,1.85)  -- (4.7,0.6);
\draw[dotted, color=revisionblue, thin] (8.8,1.85)  -- (8.8,-0.9);
\draw[dotted, color=revisionblue, thin] (12.9,1.85) -- (12.9,0.6);

\node[eventdot] at (0.6,0.6)   {};   %
\node[eventdot] at (4.7,0.6)   {};   %
\node[eventdot] at (8.8,-0.9)  {};   %
\node[eventdot] at (12.9,0.6)  {};   %

\draw[flowarrow] (4.7,0.5) -- (8.8,-0.8)
    node[midway, sloped, align=center, above=-1pt, font=\scriptsize, color=revisionblue]
        {submit PR};
\draw[flowarrow] (8.8,-0.8) -- (12.9,0.5)
    node[midway, sloped, align=center, above=-1pt, font=\scriptsize, color=revisionblue]
        {pull merged commit};

\node[anchor=south, text width=2.6cm, align=center, font=\footnotesize] at (0.6,0.75)
    {Last fork commit\\authored\\\textit{(source fork)}};
\node[anchor=south, text width=2.6cm, align=center, font=\footnotesize] at (4.7,0.75)
    {PR creation\\(same PR)\\\textit{(source fork)}};
\node[anchor=north, text width=2.6cm, align=center, font=\footnotesize] at (8.8,-1.05)
    {PR merged or closed\\\textit{(origin)}};
\node[anchor=south, text width=2.6cm, align=center, font=\footnotesize] at (12.9,0.75)
    {Another fork pulls\\the merged commit\\\textit{(another fork)}};

\draw[dashed, color=revisionblue!40] (8.8,2.95) -- (8.8,-0.9);
\end{tikzpicture}
\caption{\revision{Synchronization timeline across the three lag phases. The dashed vertical separator at PR merged or closed marks the directional boundary between Origin--Fork and Fork--Origin flows.}}
\label{fig:lag-timeline}
\end{figure}
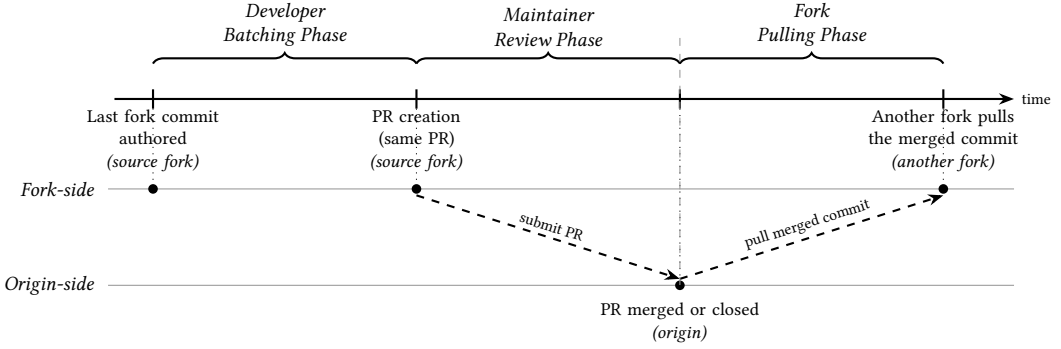
}%

\textbf{Lag Determination.} Rather than treating raw durations as ``delay'' in absolute terms, we define \textit{lag} relative to the empirical baseline of each stage: for every time interval category (e.g., Developer Batching), we compute the average duration across all observations in that interval for the corresponding stage (a single PR for the two per-PR phases, a single commit-fork pair for Fork Pulling), and we label an observation as exhibiting lag only when its duration exceeds this interval-specific average. This normalization controls for temporal shifts in project activity and review workload, and ensures that ``lag'' reflects unusually slow progression compared to contemporaneous practice.

\begin{table}[t]
\centering
\caption{Overview of synchronization lag characteristics.
Lag is reported only when the corresponding duration exceeds the empirical average for that stage.}
\label{tab:rq3-lag-overview}
\small
\begin{tabular}{lrrrrrrr}
\toprule
\textbf{Lag Type} &
\textbf{\#PRs/Forks} &
\textbf{Prevalence} &
\textbf{Median} &
\textbf{Mean} &
\textbf{P75} &
\textbf{P90} &
\textbf{Max} \\
& & & \multicolumn{5}{c}{\textbf{Delay (days)}} \\
\midrule

\multicolumn{8}{l}{\textbf{Fork$\rightarrow$Origin (PR-level, $n=158{,}485$)}} \\
\midrule
Submission Batching
& 14{,}428 & 9.1\%
& 32.0 & 120.6 & 107.3 & 311.6 & 17{,}761 \\

Review Latency
& 21{,}756 & 13.7\%
& 46.4 & 137.1 & 131.9 & 368.3 & 4{,}228 \\
\midrule

\multicolumn{8}{l}{\textbf{Origin$\rightarrow$Fork (fork-level, $n=3{,}820$)}} \\
\midrule
Fork Pulling
& 93{,}972 & 59.3\%
& 6.8 & 11.5 & 15.6 & 28.9 & 714 \\
\bottomrule
\end{tabular}
\label{tab:lag}
\end{table}

\begin{figure}[t]
    \centering
    \includegraphics[width=\linewidth]{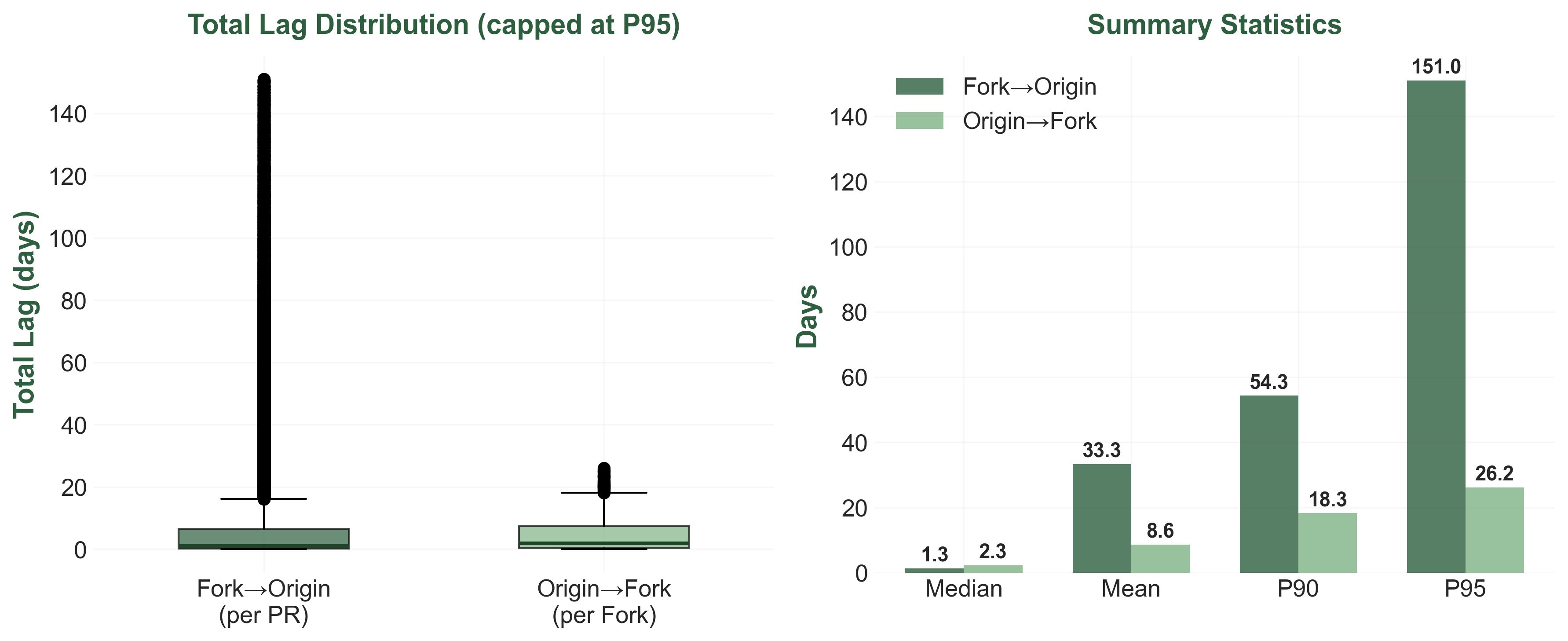}
    \caption{Comparison of total synchronization lag by direction. Fork$\rightarrow$Origin (developer batching + maintainer review per PR) vs.\ Origin$\rightarrow$Fork (average pulling lag per fork).}
    \label{fig:rq3-comparison}
\end{figure}
\subsubsection{RQ2.1: Synchronization Lag in Fork$\leftrightarrow$Origin Flows}
\label{sec:rq2-flows}

Table~\ref{tab:lag} summarizes the prevalence and severity of each lag component, while Figure~\ref{fig:rq3-comparison} contrasts the \emph{total} synchronization delay.

Table~\ref{tab:rq3-lag-overview} indicates that lag is not rare and is unevenly distributed across stages.
\revision{For fork$\rightarrow$origin PRs in our dataset, \textbf{Review Latency Lag} (13.7\%) and \textbf{Submission Batching Lag} (9.1\%) occur at broadly comparable rates; we report these descriptively and do not claim one is a more prevalent bottleneck than the other.
Moreover, \textbf{Fork Pulling Lag} is the most prevalent signal (59.3\%), reflecting both the much higher volume of downstream pulling activities compared to selective upstream PR submissions, and the fact that a large fraction of forks operate under non-trivial origin--fork desynchronization.
In terms of severity (among lagging cases), batching and review lags exhibit pronounced long tails, with high upper quantiles and extreme maxima, consistent with occasional but substantial slowdowns that dominate average delay. }

\textbf{Activity does not prevent drift:}
Surprisingly, pulling lag is not confined to inactive forks.
We observe a positive association between pull activity and pulling lag ($\rho=0.508$, $p<0.001$): forks with more pull operations tend to have higher average delays.
Concretely, we identify 414 forks with more than 1{,}000 pull operations yet more than 10 days of average pulling lag, demonstrating that high synchronization activity does not guarantee low drift.
This suggests that drift can persist even under frequent integration attempts, likely due to bursty update patterns, delayed adoption of specific upstream commits, or selective pulling strategies.

\textbf{Directional contrast in total lag:}
Figure~\ref{fig:rq3-comparison} compares direction-specific lag distributions. Although medians are comparable, fork$\rightarrow$origin synchronization exhibits a substantially heavier tail than origin$\rightarrow$fork pulling, revealing a \textit{propagation asymmetry}: even when upstream code successfully merges, it may take weeks to propagate back through the fork ecosystem.
This has direct security implications---patches merged into origins may remain unapplied in downstream forks, leaving the software supply chain exposed.

\textbf{Commit-level lifecycle confirms fork pulling as the bottleneck:}
To validate these PR-level findings, we traced the end-to-end lifecycle of 33,767 individual commits from creation through upstream merge to adoption by the last synchronized fork. The median total lifecycle is 3.17 days (90th percentile: 50.48 days). Critically, the Fork Pulling Phase accounts for \textbf{72.9\%} of the entire life cycle, despite the Developer Batching and Maintainer Review Phases involving explicit human coordination. Moreover, 38.8\% of commits experience pulling lag, nearly 3$\times$ higher than the 13.7\% review lag, confirming that fork pulling is the most consistently delayed phase in the pipeline.

\mybox{\textbf{Finding 3:} Synchronization delay is pervasive and structurally uneven. The Fork Pulling Phase accounts for 72.9\% of end-to-end delay, with 38.8\% of commits experiencing pulling lag—nearly 3$\times$ the 13.7\% affected by maintainer review. Fork pulling is the dominant bottleneck, creating persistent vulnerability windows.}

\begin{figure}[t]
    \centering
    \includegraphics[width=\linewidth]{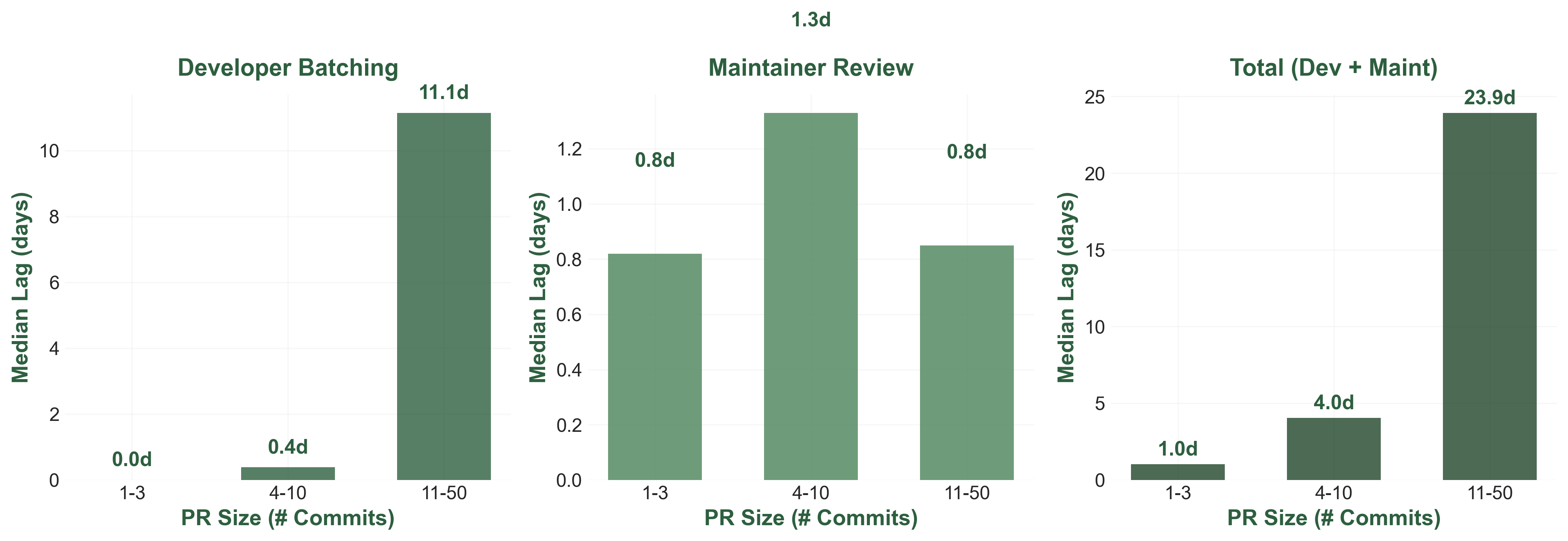}
    \caption{Fork$\rightarrow$Origin: Median lag by PR size. Developer batching lag increases substantially with PR size, while maintainer review lag remains relatively constant.}
    \label{fig:f2o_size}
\end{figure}

\textbf{PR size drives batching lag:}
Figure~\ref{fig:f2o_size} shows batching-lag prevalence rises from 4.3\% for single-commit PRs to 57.2\% for PRs with 11--30 commits (13$\times$ increase), while review latency shows no such correlation.
This implies that large PRs act as a temporal filter---disproportionately delayed before submission, reducing the likelihood of timely upstream exposure.
Because batching lag is predictable from PR size, tools can intervene early by recommending PR decomposition and prompting earlier submission of partial units.

\subsection{RQ3: Pull Request Categorization}

RQ1 and RQ2 established that synchronization may be either delayed or incomplete.
However, these findings do not explain what kinds of changes successfully synchronize and why synchronization attempts fail.
Understanding these patterns is essential for designing effective interventions.
In this RQ, we therefore categorize the intent of merged PRs to identify which contribution types propagate successfully, and analyze rejected PRs to reveal the root causes of synchronization failure.

\subsubsection{RQ3.1: Categorization of Intentions}
To analyze cross-repository contributions from forks to origin repositories, we collected \textbf{158,482} closed and merged PRs submitted by forks to their corresponding origin repositories.
To prioritize PRs that are more likely to reflect meaningful and sustained parallel development, we rank PRs in descending order based on the popularity of the contributing forks, measured by their star counts, which have been widely used as a proxy for project maturity and community interest.
From this ranked list, we selected the top \textbf{30,000} PRs for in-depth analysis to match our analysis capacity.
These PRs were further examined to infer their functional intent using a hybrid card sorting approach~\cite{hybridcardsorting,fincher2005cardsorting} with the aid of an LLM and human consolidation to dynamically derive the categorizations.

We first derived an initial set of high-level intent categories from established taxonomies of software changes and pull-request studies, including bug fixing, feature addition, refactoring, documentation, and performance optimization according to related commit classification work~\cite{mockus2000commits,hindle2008commits,gousios2014exploratory}.
These categories reflect commonly accepted dimensions of developer intent in collaborative software evolution and provide a stable conceptual starting point.

Next, we employed an LLM, DeepSeek-R3.2~\cite{deepseek}, to support the systematic classification of merged PRs by intent, following recent work demonstrating the effectiveness of LLMs for software engineering tasks~\cite{hou2024llmse}. 
For each PR, the LLM was provided with multiple information sources, including the PR title, description, commit messages, and code diffs, and was prompted to assign the PR to the most appropriate intent category. 
In addition, the model was instructed to explicitly flag PRs whose changes could not be adequately captured by any existing category.

We constructed the intent taxonomy using a hybrid card-sorting process that combines both inductive and deductive reasoning. 
Specifically, we began with a small set of high-level, pre-defined categories derived from prior studies on code change classification and software maintenance activities.
Using a randomly sampled subset of PRs, we then performed an open coding phase, in which the LLM proposed fine-grained intent labels based on observed change patterns without being restricted to the initial taxonomy.
These emergent labels were subsequently reviewed and consolidated into candidate categories.

Lastly, we applied the evolving taxonomy to the full PR dataset. 
Whenever the LLM consistently identified PRs that did not fit well into existing categories, we revisited the taxonomy and refined it through category splitting or merging, guided by semantic coherence and empirical frequency.
This iterative refinement continued until the taxonomy stabilized and no recurrent uncategorizable PRs were observed.

The resulting taxonomy has 37 categories and satisfies three design criteria: (i) categories are mutually exclusive, such that each PR is assigned to exactly one category; (ii) categories are collectively exhaustive with respect to the analyzed PRs; and (iii) category definitions are interpretable at the individual PR level.
Each PR was ultimately assigned to the single category that best reflects its dominant intent.
The full prompting protocol and taxonomy definitions are publicly available on our project website~\cite{dataset}.

\subsubsection{Categories of Intentions}
Our analysis reveals the distribution of the most frequent PR categories.
Bug fixes dominate fork-to-origin contributions, accounting for \textbf{36.42\%} (10,927 PRs) of all analyzed PRs.
This is followed by feature additions (\textbf{14.15\%}), code refactoring (\textbf{8.63\%}), documentation updates (\textbf{5.79\%}), and performance optimizations (\textbf{5.27\%}).
Together, the top five categories constitute over \textbf{70\%} of all PRs, indicating that the majority of fork contributions focus on maintenance and quality improvement rather than novel functionality.

Beyond these dominant categories, the remaining PRs form a long tail of diverse contribution types, including platform-specific modifications, dependency upgrades, configuration changes, and localization.
Although each of these categories individually represents a small fraction of PRs, their collective presence highlights the heterogeneous nature of parallel development across forks.

\begin{figure}[t!]
	\centering
\includegraphics[width=0.9\linewidth]{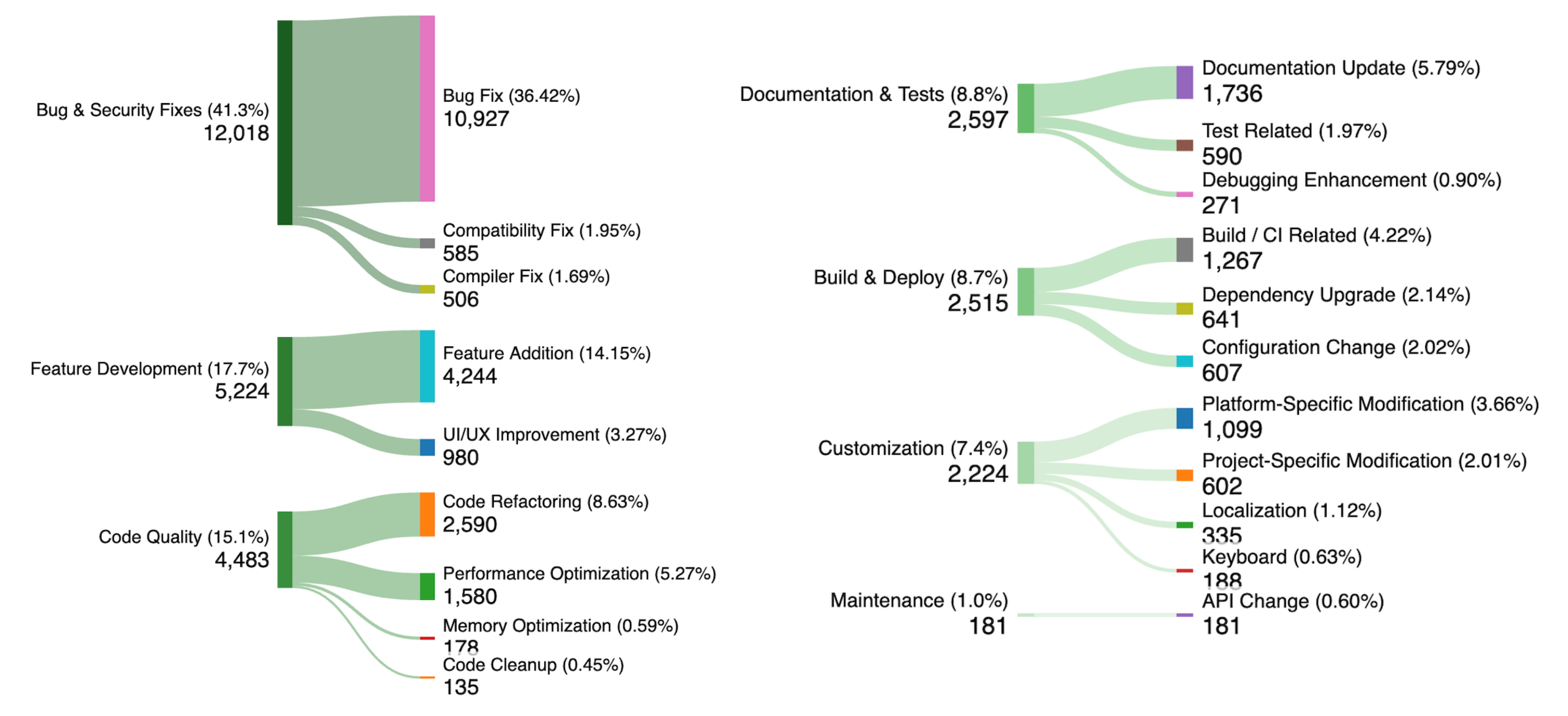}
	\caption{Thematic Grouping of PR Intention Categories}
	\label{fig:sankey}
\end{figure}

For analytical clarity, we further grouped the 37 fine-grained categories into higher-level thematic clusters as in Figure~\ref{fig:sankey}. For clarity, the categories with fewer than 100 PRs were omitted.
Bug and security-related fixes—including bug fixes, compatibility fixes, compiler fixes, and safety enhancements—constitute approximately \textbf{41.99\%} of all PRs.
Feature development accounts for \textbf{17.76\%}, while code quality improvements (e.g., refactoring and optimization) represent \textbf{9.59\%}.
Documentation and testing-related changes comprise \textbf{8.88\%}, and build or deployment-related PRs account for \textbf{8.50\%}.
Finally, customization-oriented changes, such as platform- or project-specific modifications, represent \textbf{7.42\%} of PRs.

\mybox{
\textbf{Finding 4:}
Fork-to-origin PRs are primarily maintenance-driven: bug fixes alone account for \textbf{36\%} of contributions, and bug/security-related fixes collectively represent \textbf{42\%}. Over \textbf{70\%} of PRs focus on maintenance and quality improvement---including refactoring, optimization, compatibility, and documentation---rather than new features (\textbf{17.8\%}). This concentration suggests that forks serve as a significant source of corrective and stabilizing contributions to upstream projects.
}

\subsubsection{RQ3.2: What Gets Rejected and Why}
\label{sec:rq2-rejection-causes}

RQ1 revealed a synchronization paradox: although approximately 90\% of submitted PRs are merged, only a small fraction of fork commits ever appear in PRs. To explain what prevents synchronization when it is attempted, we analyze a sample of 5,000 rejected fork$\rightarrow$origin PRs and examine the reasons they fail to synchronize. 
As discussed previously, we categorize rejection reasons using the same LLM-assisted hybrid card-sorting procedure.
Unlike the intent analysis for merged PRs, this classification focuses on rejection rationales inferred from PR discussions and review comments, and allows multiple reasons to be assigned to a single PR.

Our analysis reveals the distribution of rejection categories. Contrary to common assumptions, rejection is \emph{not} primarily driven by technical incorrectness. Only \textbf{17.3\%} involve any technical concerns. Instead, rejections concentrate in three dominant categories.

First, \emph{superseded or duplicate PRs} account for \textbf{24.6\%} of all rejections. These PRs implement functionality that has already been merged through another contribution, indicating that multiple forks often pursue similar solutions independently. This pattern reflects the parallel development observed in RQ1: when synchronization is delayed or uncoordinated, redundant solutions emerge.

Second, \emph{process issues} account for \textbf{24.1\%} of rejections. These PRs are typically technically plausible but violate contribution norms, such as submitting development branches instead of curated changes, including excessive or noisy commit histories, or lacking required formatting. Notably, these PRs are substantially larger than other rejected PRs in terms of the number of commits (mean 47.3 commits versus 11.4 for duplicates), indicating that process overhead is a significant barrier to synchronization rather than code quality.

Third, \emph{maintainer policy decisions} explain \textbf{16.7\%} of rejections. These PRs are actively reviewed (98.8\% have maintainer comments) but rejected due to misalignment with project direction, architectural vision, or roadmap priorities. Unlike process-related rejections, these decisions represent structural limits to synchronization that cannot be easily resolved through revision.

Together, these three categories account for \textbf{65.4\%} of all rejected PRs. The remaining rejections are distributed across insufficient information (11.7\%), code quality issues (10.5\%), project fit mismatch (6.2\%), and minor categories such as incorrect submission targets or communication failures. Rejection patterns are also highly concentrated: the top ten origin repositories account for \textbf{50.8\%} of all rejected PRs, suggesting that synchronization barriers are strongly shaped by project-specific norms rather than universal rules.

\mybox{\textbf{Finding 5:} PR rejection is rarely caused by technical incorrectness---only \textbf{17.3\%} of rejected PRs involve any technical concerns. Instead, over \textbf{65\%} of rejections stem from superseded contributions (24.6\%), process violations (24.1\%), or maintainer policy decisions (16.7\%), indicating that synchronization failures are driven by coordination overhead and governance constraints rather than code quality.}

\subsection{RQ4: Synchronizable Commits Mining}
\label{sec:rq4-syncability}
RQ1--RQ3 show that synchronization is selective and often blocked by non-technical barriers.
RQ4 shifts focus from PRs to fork-local commits, asking how much divergence could be avoided by identifying sync-ready commits that remain local. \revision{The platform's design point is visibility: synchronization gaps are typically invisible to downstream users and maintainers by default, and surfacing them as optional recommendations allows informed decisions without imposing automated cross-repository pushes.}

Concretely, we focus on fork-local unique commits that satisfy two principles mainly derived from previous conclusions.
First, they are \textbf{sync-worthy}: the change is broadly beneficial to other repositories (e.g., bug fixes, compatibility improvements, performance optimizations) and is not merely fork-specific new functionality tied to a local codebase.
Second, they are \textbf{sync-eligible}: the change is technically portable to the target repository (or peer forks) and does not violate repository-specific constraints such as project scope, contribution rules, or architectural invariants.
These criteria motivate our staged syncability assessment pipeline, which operationalizes technical portability through conservative checks and usefulness verification.

\textbf{Design goals:}
Our assessment targets technical mergeability rather than social acceptance.
Accordingly, the pipeline is (i) conservative: false positives are minimized; (ii) stage-gated: later checks run only if earlier checks pass; (iii) decision-based: binary pass/fail rather than weighted scoring.
To increase robustness against fork divergence, we compare candidates not only against the origin repository but also against other forks in the family, enabling multi-repository evidence for file availability and diff-context compatibility. A candidate is labeled \texttt{Synchronizable} if it passes the following stages:

\textbf{Stage 0 Candidate extraction and prioritization.}
We extract all native commits with metadata and diffs, and identify the test files by keywords to treat them separately.
The following pipeline prioritizes commits that include test files, as these tend to be better packaged and easier to validate.

\textbf{Stage 1 Criteria of Sync-worthy.}
\begin{enumerate}[leftmargin=9pt]
    \item \textbf{Large-scale refactoring filter} rejects commits that touch too many files (default threshold: $\geq 50$ files). 
Based on the findings from RQ3, we first sort candidates by the number of code files modified (ascending) and apply this filter because large, cross-cutting refactorings are typically difficult to port across repositories: they require substantial review effort, are more likely to conflict with repository-specific evolution, and often demand non-trivial manual conflict resolution and re-validation. 
    \item \textbf{Dependency-only update filter} rejects commits whose modified files are exclusively dependency or third-party management (e.g., build manifests, lock files, and \texttt{vendor/} or \texttt{third\_party/} directories). 
We apply this filter because dependency configurations are often \emph{project- and environment-specific}: forks may pin different versions, enable different feature flags, select different build options, making such changes unlikely to be directly transferable across repositories.
    \item \textbf{LLM-based usefulness screening}: given the commit message and a structured diff summary, an LLM predicts whether the change is broadly beneficial beyond the local fork.
This screening is grounded in the empirical evidence from RQ3: the intent taxonomy of merged PRs identifies the change types that are generally valued beyond the current fork (e.g., bug fixes, compatibility fixes, refactoring/quality improvements, and performance optimizations).
As long as the commit falls into the beneficial category, it passes the screening.
Note that the usefulness of certain commits varies across repository context, and it is very hard for non-maintainers to accurately determine whether a commit is useful for a given repository without understanding the whole code context. Therefore, we do not take the screening of context-dependent usefulness into account.
\end{enumerate}

\textbf{Stage 2 Criteria of Sync-eligible.}
Stage~2 performs a conservative, stepwise assessment of whether a fork-local commit can be ported to a target repository (the origin and/or peer forks).
We follow the taxonomy of patch-porting difficulty~\cite{yang2023enhancing,shariffdeen2021automated}: Type-I (direct application, no adaptation needed), Type-II (patch-location changes due to file renaming or restructuring), Type-III (symbol resolution and dependency importing), and Type-IV (significant code transformations and module-level modifications).
We conservatively operationalize sync-eligibility as covering only Type-I--II cases, and defer Type-III--IV (which requires substantial semantic adaptation and is empirically harder to automate) to future work.

\begin{enumerate}[leftmargin=9pt]
    \item \textbf{LLM-assisted Guideline compliance screening:}
    Before attempting technical porting, we use an LLM to assess whether the change is likely to violate the target repository's contribution constraints (e.g., required formatting, test expectations, and scope rules) by jointly examining \texttt{CONTRIBUTING.md}/CI guidance when available, the commit message, and a structured diff summary.
    This step reduces unnecessary commits that might be predictably rejected or impose avoidable review burden.

   \item \textbf{Patch-location \& context feasibility (Type-I--II).}
We first verify that all modified file paths exist in the latest target repository snapshot, and then validate that each diff hunk's pre-change context is present for Type-I--II porting. Specifically, the context of all hunks must match the target files under normalized namespace, approximating the minimal feasibility condition for Type-I--II porting: a valid landing location and a recognizable pre-change context in the target codebase.
If file paths are missing or hunk contexts cannot be matched, applying the change would likely require more intrusive re-targeting, which is characteristic of Type-III--IV or higher-complexity porting and is conservatively excluded in our study.

\end{enumerate}

\subsubsection{Effectiveness of Locating Synchronizable Candidates.}
\label{sec:rq4-effect}

Applying the pipeline to 3{,}820 forks yielded 5.4 million native commits.
Stage~0 prioritized commits by ranking them in ascending order of non-test file count, retaining the top 0.5 million commits that include test files. This threshold was chosen as a practical scope for a proof-of-concept evaluation; the pipeline itself is designed to scale to the full corpus in future work.
Technical eligibility checks in Stages~1--2 further reduce candidates to 24{,}116 pairs, removing commits incompatible with target repositories.
Domain filtering excludes 2{,}087 pairs touching only templates or tests and 5{,}320 pairs without source files, leaving 16{,}709 pairs.
Finally, intent screening restricts to sync-worthy categories (Bug Fix, Security Fix, Code Quality), yielding the final \textbf{12{,}284} sync-ready pairs, resulting in 2.45\% of the sampled 0.5 million.

\textbf{Pull request submission to fork families.}
Our tool identified \textbf{12,284} sync-ready commit--repository pairs.
To validate practical value, we have submitted \textbf{73} pull requests to target repositories (origins or peer forks), selecting candidates where the change is expected to be useful and acceptable.
Scaling beyond this sample was infeasible: GitHub's anti-abuse mechanisms flagged and restricted our accounts after detecting automated cross-repository PR activity.
As of this writing, 3 PRs have been merged while the rest remain under review or discussion.
The monitoring platform is currently deployed and actively tracking synchronization opportunities across the collected fork families in our dataset, with new sync-ready candidates detected within hours of commit creation.
To enable broader adoption, we plan to release the platform publicly, allowing developers to opt in and receive sync recommendations for their own fork families~\cite{dataset}.

\revision{\textbf{Supply-chain risk and maintainer control.}
Automatically pushing fork-derived changes (as bulk PRs or notifications) to maintainers or to peer forks would invite supply-chain risk and is the scenario that GitHub's anti-abuse mechanisms guard against. The platform intentionally avoids this: candidate sync-worthy commits are surfaced as optional recommendations that maintainers see only when they choose to look. The platform thereby improves the exposure of syncable commits, including stale security fixes that failed to propagate upstream.}

\revision{\textbf{Operational cost.} The platform's operational overhead is modest. For the final security-commit pipeline, LLM API cost is \$0.01--0.04 per commit, and end-to-end processing time is 2.2--5.4 seconds per commit. The full one-time experiments, including preliminary testing, cost approximately \$350.}

\textbf{Security validation of syncable commits.}
To assess whether unsynced commits expose target repositories to concrete security risks, we manually validated \textbf{153} sync-ready security-related commit--repository pairs from our dataset, covering ecosystems such as Bitcoin, Ceph, OpenWrt/LEDE, ClickHouse, Dogecoin, and others.
For each pair, we fetched the commit from GitHub, identified the security-relevant change, and then checked whether the target repository still contained the vulnerable or outdated code.
Of the 153 pairs, \textbf{83} were manually reviewed and confirmed as potential 1-day vulnerabilities: the target repository had not applied the fix and remained exposed.
These span diverse security categories including patching with known CVEs (e.g., Dropbear SSH, glibc, miniupnpc), build hardening gaps (missing compiler flags, unnecessary sample code), stale sanitizer suppressions masking real bugs, and insecure CI/Docker configurations.
We have manually produced \textbf{35} proof-of-concept so far to demonstrate these vulnerabilities and filed issues to the affected repositories to draw maintainer attention.
Continuous progress on these security findings is tracked on our monitoring platform~\cite{dataset}.

\revision{\textbf{Manual validation of sync-ready candidates.}
To verify that the LLM-based screening is reliable, two authors independently reviewed a random sample of 300 sync-ready candidates drawn from the full pipeline output, with a third resolving disagreements~\cite{seaman1999qualitative}. Of the 300 candidates, \textbf{258} (\textbf{86\%}) were judged genuinely sync-worthy with \textbf{85\%} inter-rater agreement~\cite{cohen1960kappa}, confirming that the LLM screening aligns well with human judgment. }

\section{Related Work}

\subsection{Forking Behavior and Parallel Development}

\revision{Prior fork studies establish forking motivations~\cite{dabbish2012social, jiang2017forks}, classification criteria that distinguish sustained hard forks from short-lived PR-only forks~\cite{zhou2019whatthefork, zhou2020forks}, ecosystem mappings and tooling~\cite{cosentino2018forks, ren2018forksinsight}, distributed-collaboration foundations within single projects~\cite{mockus2002apache}, and narrow security-propagation effects~\cite{lefeuvre2025oneday}. Closest to our angle, Lin et al.\ study how Linux distributions coordinate bug fixes and security patches with their upstream projects~\cite{lin2022upstream, lin2023vulnerability}, but in a curated package-level setting (Debian, Fedora) with formal integrator roles. We instead study general GitHub fork families, including independently evolving and unmaintained forks, at the commit level, with bidirectional synchronization, lag decomposition, rejection-barrier analysis, and sync-ready missed-opportunity mining as the focus. None of these prior works analyze bidirectional commit synchronization or quantify technical lag across fork families.}

\subsection{Pull Request Studies}

\revision{The pull-based development model is well-studied. Rigby et al.~\cite{rigby2008peer} characterize open-source peer review as early, frequent reviews of small, independent contributions; our size-effect on batching lag corroborates this in the fork$\to$origin setting. Gousios et al.~\cite{gousios2014exploratory, gousios2015integrator} explore pull-based development and the integrator's perspective. Zhang et al.~\cite{zhang2022prdecision, zhang2022prlatency} analyze merge-decision factors and PR latency, and Bacchelli and Bird~\cite{bacchelli2013codereview} study modern code-review practice. Prior studies of patches, pull requests, and modern code review~\cite{jiang2013patches, tao2014writing, egelman2020feelings, wang2019abandoned, kononenko2018studying, zhang2022prdecision, zhang2023compatible} characterize merge decisions, latency, rejection, abandonment, and review friction, with a mix of technical, process, social, and project-context factors. None of these address fork-family synchronization specifically.}

\subsection{Patch Porting and Commit Migration}

Another line of work examines how patches and commits are reused, duplicated, or ported across related codebases.
Li et al.~\cite{bissyande2013patchporting} investigated patch porting practices in the Linux kernel ecosystem and showed that patches are frequently backported across branches with varying strategies and delays.
Nguyen et al.~\cite{nguyen2013dupfix} demonstrated that similar patches often appear in multiple locations due to cherry-picking~\cite{bunyakiati2017cherry} or manual migrations, and code clone detection techniques~\cite{roy2007clonesurvey,sajnani2016sourcerercc} have been applied to identify such duplicated changes.
Pan et al.~\cite{pan2024automating} proposed PPatHF, an LLM-based approach for automating zero-shot patch porting across hard forks, demonstrating that 42.3\% of patches from Vim could be automatically ported to Neovim, highlighting both the feasibility and the remaining difficulty of cross-fork patch migration.
Merge conflicts~\cite{brun2011proactive,ghiotto2018merge} further complicate synchronization as codebases diverge. Hu et al.~\cite{hu2024empirical} studied the patch propagation in the Golong ecosystem.
However, prior work focuses on branches within a single project or on specific fork pairs; the broader fork ecosystem on GitHub, where commit synchronization spans over parallel development, remains far less understood.

\section{Discussion}
\subsection{Threats to Validity}
\label{sec:threats}

\subsubsection{External Validity.}
\textbf{Platform scope (GitHub only).}
Our study focuses on GitHub repositories to enable consistent cross-repository commit tracking and to leverage rich contribution metadata (e.g., PR discussions, merge outcomes, and review timelines)~\cite{kalliamvakou2014perils,bird2009git}. This choice may bias findings toward GitHub-specific development practices. Nevertheless, GitHub hosts a diverse range of projects and provides PR-centric signals that are difficult to recover from broader archival sources (e.g., Software Heritage) that lack comparable contribution and review process data.

\textbf{Fork-family construction.}
We identify actively maintained forks using observable signals such as stars and commit activity. These indicators may misclassify some forks (e.g., actively used but low-star forks, or periodically maintained forks with sparse commits). However, given the absence of reliable ground truth for "active maintenance" at scale, following the prior work, these signals provide a practical and reproducible approximation for deriving fork families.

\textbf{Temporal bias.}
Our analysis is based on a snapshot of repository data collected at a fixed point. Synchronization activity that occurred after our data collection window, or commits that were synchronized shortly before our snapshot but not yet propagated to all forks, may be missed. This limitation is inherent to empirical studies of evolving repositories; we mitigate it by analyzing long observation windows and focusing on structural patterns rather than point-in-time counts.

\revision{\textbf{Selection bias toward popular repositories.}
Our 2,000 star-ranked origins are concentrated in the popular end of the GitHub distribution (rationale in Section~3), so synchronization patterns measured here may not transfer to long-tail projects with weaker maintainer engagement and smaller fork ecosystems. Headline numbers (90\% merge rate, 65\% governance-dominated rejection mix, 72.9\% pulling-lag share) could shift for the unselected projects, and quantitative magnitudes such as CER values and lag durations should be read as estimates conditional on the popular projects. We mitigate this in two ways: we report medians and long-tail distributions rather than only means; we ground qualitative findings in mechanisms that do not depend on popularity.}

\subsubsection{Internal Validity.}
\textbf{LLM-based categorization and taxonomy refinement.}
Our intent and rejection-reason analyses rely on LLM-assisted hybrid card sorting, which may introduce misclassification. We mitigate this by manually validating 300 random sync-ready commit--repository pairs (two independent reviewers, a third resolving disagreements), yielding \textbf{85\%} inter-rater agreement and \textbf{86\%} confirmed sync-worthy (Section~\ref{sec:rq4-effect}), and by using a validator-role setup to reduce systematic LLM errors. Our policy-compatibility screening likewise relies on LLM interpretation of repository guidelines, which we partially validate against real PR submission behavior.

\revision{\textbf{Subjectivity of manual sync-worthiness validation.} The human review reflects author judgment rather than upstream-maintainer acceptance, and the sample ($\approx$2.4\% of 12{,}284 candidates) bounds the precision of the estimate. However, considering the difficulty in obtaining the ground truth from upstream maintainers, and manual efforts, we only sampled 300 candidates for validation,
and release the annotated sample for independent audit.}

\textbf{Syncability pipeline assumptions and false negatives.}
Our syncability assessment is intentionally conservative and may exclude commits that are technically portable but require non-trivial adaptation, increasing false negatives. We prioritize high-confidence, sync-worthy candidates over noisy recommendations, using prioritization rather than hard thresholds and reserving hard assumptions for technical eligibility checks. Our current operationalization targets Type-I--II porting; more complex cases remain hard to automate robustly and are left for future work.

\textbf{Limited PR submission validation.}
We validated the pipeline by submitting 73 PRs to target repositories. Although small relative to the 12,284 sync-ready pairs (0.6\%), scaling up was infeasible: GitHub's anti-abuse mechanisms flagged and restricted our accounts after detecting automated cross-repository PR activity, even with authenticated identities, reflecting a tension between large-scale synchronization research and anti-spam policies.
To address this, we progressively release our monitoring platform publicly so developers can opt in and receive sync recommendations for their own fork families rather than unsolicited PRs.

\subsection{Discussion: Lessons Learned}
\label{sec:discussion}

\textbf{Lesson 1: Effective synchronization tooling must be governance- and value-aware, not merely technically feasible.}
Our rejection analysis (Finding~5, Section~\ref{sec:rq2-rejection-causes}) shows that over 65\% of rejected PRs are blocked by process violations, maintainer policy decisions, or redundancy rather than technical defects, while RQ4 (Section~\ref{sec:rq4-syncability}) reveals that many fork-local commits are mechanically portable yet offer limited value or conflict with target-repo constraints. Together, these findings imply that synchronization tools must go beyond patch applicability to jointly consider governance compatibility and semantic value. Our monitoring platform operationalizes this insight by filtering candidates through technical portability, usefulness grounded in empirically observed beneficial contribution types (Finding~4), and policy compatibility derived from repository guidelines. For practitioners, this reframes synchronization risk: investing solely in better patch application is insufficient if submissions systematically violate norms. For the research community, our evidence base makes implicit governance barriers explicit, motivating policy-awareness as a first-class design criterion for synchronization tools.

\revision{\textbf{Lesson 2: Rebased or amended PR commits make SHA-only exposure metrics conservative.} Our classification of 532{,}070 commits in 158{,}472 fork$\to$origin PRs (RQ1.2, Section~\ref{sec:rq1-synchronization-paradox}) shows that 32.73\% do not match any fork SHA but have a diff at least 85\% similar to a fork unique commit, indicating a rewritten identity from rebase or amend. At the PR level, 33.72\% of PRs consist entirely of derivative commits. These derivative commits are not incidental noise; they reflect normal contribution preparation, such as rebasing on the latest upstream state or amending commits before opening a PR. Treating them as unexposed would misclassify synchronized changes as missing simply because the commit identifier changed. We therefore report a history-rewriting-aware CER alongside the exact-SHA baseline, separating the conservative lower bound from the content-aware synchronization estimate.

\textbf{Lesson 3: Synchronization delay is dominated by fork pulling, not by PR review.} Our commit-level lifecycle trace of 33{,}767 commits (Finding~3, RQ2, Section~\ref{sec:rq2-lag}) shows that the Fork Pulling Phase accounts for 72.9\% of end-to-end commit lifecycle delay, whereas developer batching and maintainer review together account for the remaining 27.1\%. The prevalence of pulling lag (38.8\% of commits) is roughly 3$\times$ the 13.7\% prevalence of maintainer review lag. The high activity does not protect against drift: among 414 forks with more than 1{,}000 pull operations, the average pulling lag still exceeds 10 days. Tooling should therefore prioritize accelerating post-merge propagation which is exactly what our tool aims to address.

\textbf{Lesson 4: A high PR merge rate can mask systemic under-exposure of fork-local work, a synchronization paradox invisible to PR-only metrics.} Although 90\% of submitted fork$\to$origin PRs are merged, the median fork exposes only 7.19\% of its post-fork commits through PRs at all (Finding~2, RQ1.2, Section~4.1.2), and roughly 2.9~million fork-local commits across our dataset never appear in any PR to the origin: a project can simultaneously sustain an excellent merge rate and a deeply unsynchronized fork ecosystem. The actionable consequence is that PR acceptance is the wrong primary signal for ecosystem-wide synchronization health. Tools and metrics targeting fork-family synchronization need the fork-level visibility that quantifies the unexposed commits and their syncability, rather than relying on PR acceptance rates alone. This is precisely the design point our monitoring platform operationalizes.

\textbf{Lesson 5: Fork$\to$origin rejections are mainly governance-driven, not technical.} Prior patch, PR, and code-review studies~\cite{jiang2013patches, tao2014writing, egelman2020feelings, wang2019abandoned, kononenko2018studying, zhang2022prdecision} show that merge outcomes reflect technical, process, social, and project-context factors. In our fork$\to$origin setting, technical incorrectness accounts for only 1\% of rejections, while process violations, maintainer-policy decisions, and superseded contributions account for over 65\% (Finding~5, RQ3, Section~4.3). These failures often occur when upstream has already addressed the issue, the fork diverges from upstream scope, or the submission violates project norms. Synchronization tools should therefore prioritize project-specific PR preparation and avoid resubmitting policy-blocked or superseded changes.}

\section{Conclusion}
\label{sec:conclusion}

This paper presents the first large-scale empirical study of fork synchronization across 2,000 GitHub origins, 3,820 forks, and 2.9 million fork-local commits. We find that only \revision{7.19\%} of fork commits appear in PRs despite a 90\% merge rate, that fork pulling accounts for 72.9\% of synchronization delay, and that over 65\% of PR rejections are governance-driven rather than technical. Guided by these insights, our three-stage syncability pipeline surfaces 12,284 sync-ready pairs (86\% confirmed sync-worthy). Future work includes extending to Type-III--IV patch porting and deploying the monitoring platform for community adoption.

\section*{Acknowledgments}
This research is supported by the National Research Foundation Singapore, Prime Minister's Office, Singapore, and the Cyber Security Agency under the National Cybersecurity R\&D Programme (NCRP25-P04-TAICeN) and its Campus for Research Excellence and Technological Enterprise (CREATE) programme. 
\clearpage
\section*{Data Availability}
\revision{Our anonymized dataset, analysis scripts, platform source code, and documentation describing the platform-construction process are available at \url{https://github.com/zly123987/fork-study}.}

\bibliographystyle{ACM-Reference-Format}
\bibliography{acmart}
\end{document}